\documentclass[onecolumn,amsmath,amssymb,11pt]{revtex4}

\def\gsim{ \lower .75ex \hbox{$\sim$} \llap{\raise .27ex \hbox{$>$}} }

\def\lsim{ \lower .75ex \hbox{$\sim$} \llap{\raise .27ex \hbox{$<$}} }

\usepackage{graphicx}
\usepackage{dcolumn}
\usepackage{bm}
\input epsf
\def\gsim{ \lower .75ex \hbox{$\sim$} \llap{\raise .27ex \hbox{$>$}} }
\def\lsim{ \lower .75ex \hbox{$\sim$} \llap{\raise .27ex \hbox{$<$}} }

\usepackage{graphicx}
\usepackage{dcolumn}
\usepackage{bm}

\newcommand{\nn}{\nonumber}
\newcommand{\be}{\begin{equation}}
\newcommand{\ee}{\end{equation}}
\newcommand{\bea}{\begin{eqnarray}}
\newcommand{\eea}{\end{eqnarray}}

\def\d{\delta}

\def\s{\sigma}

\def\e{\epsilon}

\def\k{\kappa}

\def\o{\omega}

\def\th{\theta}

\def\s{\sigma}

\def\z{\zeta}

\begin{document}

\title{Non-Gaussianity Generated by the Entropic Mechanism in Bouncing Cosmologies Made Simple}

\author{Jean-Luc Lehners$^a$ and Paul J. Steinhardt$^{a,b}$}
\affiliation{$^a$ Princeton Center for Theoretical Science,
Princeton University, Princeton, NJ 08544 USA \\ $^b$ Joseph
Henry Laboratories, Princeton University, Princeton, NJ 08544
USA }

\begin{abstract}
Non-gaussianity in the microwave background radiation is bound
to play a key role in giving us clues about the physics of the
very early universe. However, the associated calculations, at
second and even third order in perturbation theory, tend to be
complicated to the point of obscuring simple underlying
physical processes. In this note, we present a simple analytic
procedure for approximating the non-linearity parameters
$f_{NL}$ and $g_{NL}$ for cyclic models in which the
cosmological perturbations are generated via the entropic
mechanism. Our approach is quick, physically transparent and
agrees well with the results of numerical calculations.
\end{abstract}

\pacs{PACS number(s): 98.80.Es, 98.80.Cq, 03.70.+k}

\maketitle

Observations of the cosmic microwave background radiation are
quickly becoming detailed enough that within the next few years
we can hope to obtain highly informative limits on the
bispectrum (and perhaps even the trispectrum) of primordial
curvature perturbations \cite{Komatsu:2009kd}. In this respect
the detection/non-detection of non-gaussianity will provide a
powerful tool in discriminating between various theoretical
models for the early universe. In simple inflationary models,
the inflaton field is an almost free field, and correspondingly
the curvature perturbations that these models generate are
governed by very nearly gaussian statistics
\cite{Maldacena:2002vr}. More complicated inflationary models,
such as multi-field models, can produce pretty much any value
for the so-called ``local'' non-linearity parameters $f_{NL}$
and $g_{NL}$ (corresponding to ``squeezed'' configurations in
momentum space), which makes it difficult to predict a natural
range \cite{Lyth:2002my}. So-called DBI models in which the
inflaton possesses a non-canonical kinetic term lead to more
distinct non-gaussian signals, involving quite different
(``equilateral'') momentum configurations in their correlation
functions \cite{Alishahiha:2004eh}, and multi-field DBI models
can even produce significant contributions of both local and
equilateral type simultaneously \cite{RenauxPetel:2009sj}. In
ekpyrotic \cite{Khoury:2001wf} and cyclic models
\cite{Steinhardt:2001st}, the cosmological perturbations are
generated during a slowly contracting ekpyrotic phase with
ultra-stiff equation of state $w_{ek} \gg 1$ (see
\cite{Lehners:2008vx} for a review). Such a phase can be
modelled via scalar fields with steep negative potentials. The
steepness of the potentials implies that these scalars are
necessarily self-interacting, and this leads to natural values
of the local non-linearity parameters that are in a range that
will be accessible to near-future observations
\cite{Koyama:2007if,Buchbinder:2007at,Lehners:2007wc,Lehners:2009ja}.
Of course, this is why it is important to understand the
physics that is responsible for the non-gaussian signals well,
so that, in case of a detection, the consequences can be best
appreciated.

Here, we will focus exclusively on cyclic models in which the
cosmological perturbations are generated by the entropic
mechanism, as this is currently the best understood mechanism
for producing a nearly scale-invariant spectrum of curvature
perturbations during a contracting phase \cite{Lehners:2007ac}.
Recently, both the associated bispectrum and  trispectrum have
been calculated numerically
\cite{Lehners:2007wc,Lehners:2009ja}. These calculations are
rather involved, and do not provide many clues about the final
outcome. Hence it is desirable to develop analytic methods,
even though they might only be approximate, to understand the
physics of these calculations more thoroughly. Some headway in
this direction was made by the authors in a recent publication
dealing with the bispectrum calculation \cite{Lehners:2008my}.
In this note we present a new and much simplified approach that
can in fact be applied to the calculation of the trispectrum as
easily as to that of the bispectrum.

The models under consideration can be described by gravity
minimally coupled to two scalar fields with potentials. Nearly
scale-invariant entropy perturbations are generated first,
during a slowly contracting ekpyrotic phase that at the same
time resolves the cosmological flatness puzzle. Subsequently,
in the approach to the big crunch, the ekpyrotic potential
becomes unimportant, and the universe enters a phase dominated
by the kinetic energy of the scalar fields. During this phase,
the entropy perturbations are converted into adiabatic
curvature perturbations with the same spectrum, and these
curvature perturbations form the seeds of the large-scale
structure during the subsequent expanding phase. We will
discuss the kinetic conversion phase in more detail below.

\begin{figure}[t]
\begin{center}
\includegraphics[width=0.75\textwidth]{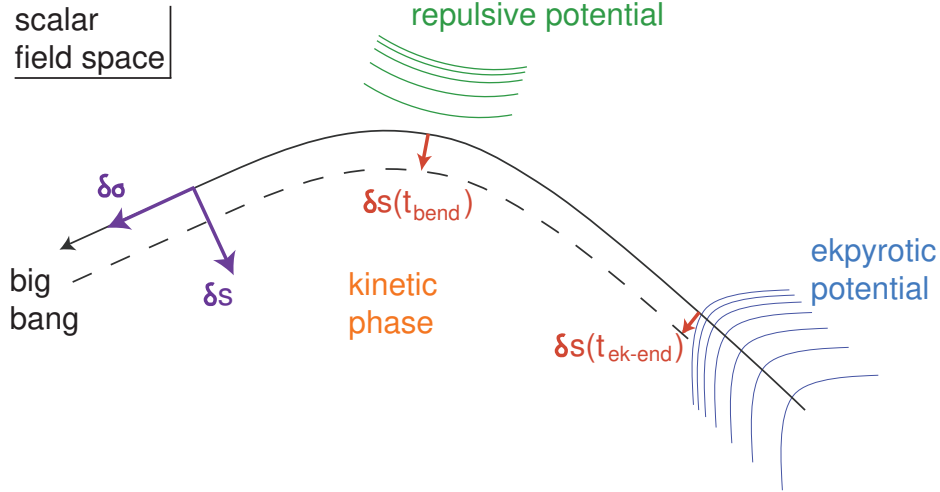}
\caption{\label{Figure1} {\small After the ekpyrotic phase, the trajectory in scalar field space enters the kinetic phase and bends - this bending is described by the existence of an effective repulsive potential (the potentials are indicated by their contour lines). A trajectory adjacent to the background evolution can be characterized by the entropy perturbation $\d s(t_{ek-end})$ at the end of the ekpyrotic phase, leading to a corresponding off-set $\d s(t_{bend}),$ or equivalently $\d V(t_{bend}),$ at the time of bending.
}}
\end{center}
\end{figure}

We adopt the following parametrization of the potential during
the ekpyrotic phase: \be V_{ek}=-V_0 e^{\sqrt{2\e}\s}[1 +\e s^2
+\frac{\k_3}{3!}\e^{3/2} s^3 +\frac{\k_4}{4!}\e^2
s^4+\cdots],\ee where we expect $\k_3,\k_4 \sim {\cal O}(1)$
and where $\e \sim {\cal O}(10^{2})$ is related to the
ekpyrotic equation of state $w_{ek}$ via $\e=3(1+w_{ek})/2.$ We
use $\s$ to denote the adiabatic direction, {\it i.e.} the
direction tangent to the scalar field space trajectory, and $s$
to denote the ``entropy'' direction, {\it i.e.} the direction
perpendicular to the background trajectory (note that the
fields $\s$ and $s$ are thus defined such that the coordinate
system they imply moves along with the background trajectory
\cite{Langlois:2006vv}); see Fig. \ref{Figure1}. The ekpyrotic
potential is tachyonic in the entropy direction, and this
instability causes the entropy perturbations to grow
\cite{Lehners:2007ac}. Moreover, this instability has the
consequence that the global structure becomes a ``phoenix''
universe, in which the universe loses most of space to black
holes at the end of each cycle, while the regions that survive
the big bang are aided by the dark energy to grow into vast new
habitable regions -- this was discussed in some detail in
\cite{Lehners:2008qe}.

During the ekpyrotic phase, it is straightforward to solve for
the entropy perturbation, with the result that
\cite{Lehners:2009ja} \be \d s= \d s_L + s_2 \d s_L^2 + s_3 \d
s_L^3, \label{entropyperturbation}\ee with the linear, gaussian
part $\d s_L$ being inversely proportional to time $t$ (defined
below in Eq. (\ref{metric})); the coefficients $s_2$ and $s_3$
are given in terms of the parameters of the potential by \bea
s_2 &=& \frac{\k_3 \sqrt{\e}}{8}, \\ s_3 &=&
(\frac{\k_4}{60}+\frac{\k_3^2}{80}-\frac{2}{5}) \e. \eea

The local non-linearity parameters $f_{NL}$ and $g_{NL}$ can be
defined via an expansion of the curvature perturbation $\zeta$
in terms of its linear, gaussian part $\zeta_L,$ \be \z = \z_L
+ \frac{3}{5} f_{NL} \z_L^2 + \frac{9}{25} g_{NL} \z_L^3. \ee
Then, it was found numerically that, for conversions lasting on
the order of one e-fold of contraction of the scale factor, the
non-linearity parameters can be well fitted by the simple
formulae  \cite{Lehners:2007wc,Lehners:2009ja}
\bea f_{NL} &\approx 12 s_2 + 5 &= \frac{3}{2} \, \k_3 \sqrt{\e} +5 \label{fNLnumerical}\\
g_{NL} &\approx 100 s_3 &= 100 \,
(\frac{\k_4}{60}+\frac{\k_3^2}{80}-\frac{2}{5}) \e.
\label{gNLnumerical} \eea
The simplicity of the end result
(given the complications of the third order perturbation
equations involved) suggests that there ought to be a more
straightforward way to obtain it. In fact, the physics of the
kinetic phase (which follows the ekpyrotic phase, and during
which the conversion takes place) is really quite simple, and
moreover, except for the fact that its initial conditions
involve the entropy perturbation $\delta s,$ the kinetic phase
has no memory of the details of the ekpyrotic phase. In
particular, only the total $\delta s$ in
(\ref{entropyperturbation}) matters, and the way we choose to
decompose it into linear, second- and third-order parts is
irrelevant at this point. This realization is the first
ingredient of our calculation.

The second is a compact and very useful expression for the
evolution of the curvature perturbation $\z$ on large scales
and in comoving gauge \cite{Lyth:2004gb,Buchbinder:2007at}: \be
\dot\z= \frac{2\bar H \d V}{\dot{\bar \s}^{2} -2 \d V},
\label{Zetadot}\ee where a dot denotes a derivative w.r.t. time
$t$, $H\equiv \dot{a}/a$ is the Hubble parameter and $a$ the
scale factor, $\d V \equiv V(t,x^i)-\bar V(t)$ and a bar
denotes a background quantity. This equation is exact in the
limit where spatial gradients can be neglected, and can thus be
expanded up to the desired order in perturbation theory if
required. First, let us present its derivation
\cite{Lyth:2004gb}: considering only very large scales, we can
write the metric as \be ds^2 = -dt^2 + a^2(t) e^{2\z(t,x^i)}
dx^i dx_i, \label{metric}\ee where all the inhomogeneities are
in $\z.$ This defines $\z$ to all orders in the long-wavelength
limit. Then the equation of continuity reads \be \dot\rho + 3
(H+\dot\z) (\rho + P) = 0, \ee where $\rho$ is the (scalar)
matter energy density, and $P$ its pressure. But we're
interested in the curvature perturbation on surfaces of uniform
energy density, so $\rho = \bar \rho$ (and hence also $H=\bar
H$). And since $\bar\rho$ satisfies $\dot{\bar\rho} + 3\bar
H(\bar\rho + \bar P)=0,$ we immediately obtain \be \dot\z=-\bar
H \frac{\d P}{\bar\rho + \bar P + \d P}. \label{eq1}\ee Now,
since we choose to consider hypersurfaces on which $\d \rho =
0,$ we obtain the relations $\d(\dot{\s}^2)=-2\d V$ and thus
$\d P = -2 \d V.$ Plugging these relations into (\ref{eq1})
then yields our desired result, Eq. (\ref{Zetadot}).

Incidentally, by expanding Eq. (\ref{Zetadot}), it is also
possible to show that it is equivalent to the third-order
equation derived in \cite{Lehners:2009ja} using the covariant
formalism. The crucial thing is to keep in mind the definitions
of the adiabatic perturbation $\delta \s$ and the entropic one
$\d s$ at higher orders, provided in \cite{Langlois:2006vv} and
\cite{Lehners:2009ja}. In particular, we have (from hereon we
drop the bar on background quantities - this should not lead to
any confusion) \bea \d V &=& V_{,s} \d s + V_{,\s} \d \s \nn \\
&& + V_{,s}[\d s^{(2)} + \frac{1}{\dot\s}\d \s \dot{\d s} +
\frac{\dot\th}{2\dot\s}(\d \s)^2]+V_{,\s}[\d
\s^{(2)}-\frac{1}{2\dot\s}\d s \dot{\d s}]\nn \\ && +
\frac{1}{2}V_{,ss}(\d s)^2 + V_{,s \s} \d s \d \s +
\frac{1}{2}V_{,\s\s} (\d \s)^2 \nn \\ && +\cdots
\label{Vexpansion}\eea Up to third order and in comoving gauge
($\d \s=\d\s^{(2)}=\d\s^{(3)}=0$) we then get \bea \d V &=& V_{,s} \d s  \nn \\
&& + V_{,s}\d s^{(2)} +V_{,\s}(-\frac{1}{2\dot\s}\d s \dot{\d
s}) + \frac{1}{2}V_{,ss}(\d s)^2 \nn \\ && +  V_{,s}\d s^{(3)}
-V_{,\s}[\frac{1}{2\dot\s}(\d s \d
s^{(2)})\dot{}+\frac{\dot\th}{6\dot\s}(\d s)^2 \dot{\d s}] +
V_{,ss}\d s \d s^{(2)} - \frac{V_{,s\s}}{2\dot\s}  (\d s)^2
\dot{\d s}+\frac{1}{6}V_{,sss}(\d s)^3,
\label{Vexpansion_comoving}\eea and by expanding
(\ref{Zetadot}) we obtain \bea \dot\z &=& \frac{2H}{\dot\s^{2}}
V_{,s} \d s \nn \\ &+& \frac{2H}{\dot\s^{2}} [V_{,s} \d
s^{(2)}-\frac{1}{2\dot\s}V_{,\s}\d s \dot{\d
s}+\frac{1}{2}V_{,ss}(\d s)^2] +  \frac{4H}{\dot\s^{4}}
V_{,s}^2 (\d s)^2 \nn \\ &+& \frac{2H}{\dot\s^{2}} [V_{,s} \d
s^{(3)}-\frac{1}{2\dot\s}V_{,\s}(\d s \d
s^{(2)})\dot{}-\frac{\dot\th}{6\dot\s^{2}}V_{,\s}(\d
s)^2\dot{\d s}+V_{,ss}\d s \d s^{(2)} -
\frac{V_{,s\s}}{2\dot\s}(\d s)^2\dot{\d
s}+\frac{1}{6}V_{,sss}(\d s)^3] \nn \\ &&+
\frac{8H}{\dot\s^{4}} [V_{,s}^2 \d s \d
s^{(2)}-\frac{1}{2\dot\s}V_{,s}V_{,\s}(\d s)^2\dot{\d s} +
\frac{1}{2}V_{,s} V_{,ss}(\d s)^3] + \frac{8H}{\dot\s^{6}}
V_{,s}^3 (\d s)^3,  \eea which agrees precisely with the
equation derived in \cite{Lehners:2009ja}. Having shown this
equivalence, we will now stick with the simple and compact form
(\ref{Zetadot}).

We are assuming that, during the kinetic phase, the trajectory
in scalar field space contains a bend, and this bend is what
causes the entropy perturbations to source the curvature
perturbations. In cyclic models embedded into heterotic
M-theory, such a bend occurs naturally because the scalar field
space contains a boundary which effectively acts as a repulsive
potential (we refer the reader to Ref.
\cite{Lehners:2006pu} for
details). However, we are simply citing this example as a
concrete realization. Our calculation applies to all cases
where there is a bend in the trajectory, although for extreme
(and unnatural) cases where the bending angle is close to $0^o$
or $180^o$ some of our approximations below might break down.
We are assuming that this bend can be described as being caused
by a monotonic repulsive potential, as depicted in Fig.
\ref{Figure1}.

Now, the third and last ingredient of our calculation is the
simple relationship between $\d V$ and $\d s$ during the
conversion process. During the ekpyrotic phase, the curvature
perturbation picks up a blue spectrum \cite{Khoury:2001zk} and
is hence completely negligible on large scales. To be precise,
since $\d V \neq 0$ during ekpyrosis, there is already some
conversion of entropy into curvature perturbations occurring at
this stage. However, this contribution is entirely negligible
compared to the subsequent conversion (see
\cite{Lehners:2008my,Lehners:2009ja} and note that since
$V_{,s}=0$ during ekpyrosis, $\d V$ starts out at subleading
order), and hence we can take $\z(t_{ek-end})\approx 0$ where
$t_{ek-end}$ denotes the time at the end of the ekpyrotic
phase, or equivalently, at the start of the kinetic phase.
Moreover, as we will see below, at the end of the conversion
process $\z$ is still significantly smaller than $\d s,$ and
hence, during the conversion process, we can take the potential
to depend only on $\d s.$ And since the repulsive potential is
monotonic, and we are interested in small departures $\d s \ll
1$ from the background trajectory, it is intuitively clear that
$\d V$ is directly proportional to $\d s$ during the bending. A
numerical calculation readily confirms this simple
relationship.

During the conversion, the effect of the repulsive potential is
to cause the entropy perturbation to behave approximately
sinusoidally, independently of the precise functional form of
the potential (this was shown analytically in
\cite{Lehners:2008my}): \be \d s \approx \cos[\o(t-t_{c})]\d
s(t_c),\ee where $t_c$ denotes the time at which the conversion
starts. Moreover, the precise value of $\d s(t_c)$ is
unimportant for the present calculation. The frequency $\o$ is
of ${\cal O}(1/\Delta t),$ where $\Delta t$ is the duration of
the conversion; the more careful analysis presented in
\cite{Lehners:2008my} leads to $\o \approx 2.5/\Delta t.$
Another useful quantity is the rate of change of the angle of
the trajectory in scalar field space \cite{Gordon:2000hv} \be
\dot\th\equiv -\frac{V_{,s}}{\dot\s} \approx \frac{1}{\Delta
t}. \ee Also, the scale factor and the scalar field velocity
along the background trajectory are rather unaffected by the
presence of the repulsive potential, so that they simply assume
the values they would in the absence of any potential \be
H=\frac{1}{3t}, \qquad \dot\s=\frac{-\sqrt{2}}{\sqrt{3}t}. \ee

As we will confirm below, during the conversion process $\d V
\ll \dot\s^2,$ so that Eq. (\ref{Zetadot}) simplifies further
to \be \dot\z \approx  \frac{2\bar H}{\dot{\bar\s}^{2}} \d V.
\label{Zetadotapprox}\ee Then, at linear order, we immediately
obtain \bea \z_L &=& \int_{bend} -\frac{2H}{\dot\s}\dot\th \d
s_L \\ &\approx& \sqrt{\frac{2}{3}} \frac{\dot\th}{\omega}
\sin(\omega \Delta t) \d s(t_c) \\ &\approx& \frac{1}{5} \delta
s(t_c). \label{zetalinear} \eea  But, as argued above, $\d s$
as a whole must behave approximately in this way during the
conversion phase, and subsequently analogous relationships hold
at higher orders too: \be \zeta^{(2)} \approx \frac{1}{5} s_2
\d s_L^2 \qquad  \zeta^{(3)} \approx \frac{1}{5} s_3 \d s_L^3.
\ee These expressions immediately allow us to calculate the
non-linearity parameters \bea f_{NL} \equiv
\frac{5}{3}\frac{\z^{(2)}}{\z_L^2} &\approx \frac{5}{3}5 s_2
&\approx 8 s_2 \\ g_{NL} \equiv
\frac{25}{9}\frac{\z^{(3)}}{\z_L^3} &\approx \frac{25}{9}5^2
s_3 &\approx 70 s_3. \eea Thus, without much work at all, and
to better accuracy than a factor of 2, we recover the
numerically (and laboriously) obtained fitting formulae in Eqs.
(\ref{fNLnumerical})-(\ref{gNLnumerical}) above.

Before discussing this result, let us briefly pause to verify
the approximation made in obtaining Eq. (\ref{Zetadotapprox}):
during the kinetic phase, we can rewrite (\ref{Zetadot}) as \be
\dot\z = \frac{t \, \d V}{1-3 t^2 \d V}. \ee The approximation
made above consists in writing $\dot\z \approx t\d V$ and this
leads to $\z \approx \frac{1}{2} t_{bend}^2 \d V(t_{bend}).$
But we know that by the end of the conversion process $\z
\approx \frac{1}{5} \d s$ and hence we find that \be
3t_{bend}^2 \d V \approx \d s \ll 1,\ee which shows that the
approximation is self-consistent and confirms the validity of
(\ref{Zetadotapprox}).

So what does our result tell us? The main point is that due to
the simplicity of the kinetic phase, the non-linearity that was
present in the entropy perturbation gets transferred
straightforwardly to the non-linearity in the curvature
perturbation. Our calculation therefore explains why there are
no significant additional constant terms in
(\ref{fNLnumerical}) or constants and $\k_3$-dependent terms in
(\ref{gNLnumerical}); {\it a priori,} there was no reason for
such terms to be absent.

Moreover, the overall magnitude of $f_{NL}$ and $g_{NL}$ is set
solely by the efficiency of the conversion process, as
expressed by the relationship between $\d s_L$ and $\z_L$ in
Eq. (\ref{zetalinear}). The fact that no additional parameter
enters into Eq. (\ref{zetalinear}) has important consequences
for observations, as it determines the scaling of the
non-linearity parameters with the equation of state parameter
$\e,$ as expressed in Eqs. (\ref{fNLnumerical}) and
(\ref{gNLnumerical}). A natural value for $\e$ would be about
$50$, so that we can expect $f_{NL}$ to be of order a few tens,
with the sign typically determined by the sign of $\k_3,$ and
$g_{NL}$ to be of order a few thousand and typically negative
in sign. These values represent the natural values predicted by
models making use of the entropic mechanism. They comfortably
fit current observational bounds \cite{Smith:2009jr,Desjacques:2009jb} while being detectable by
near-future observations. It is useful to contrast these values
with those predicted by ``new ekpyrotic'' models
\cite{Buchbinder:2007ad,Creminelli:2007aq} where the entropy
perturbations are converted into curvature perturbations
directly during the ekpyrotic phase. For this variant
conversion process, the dependence on the equation of state
$\e$ is more pronounced, with $f_{NL} \propto \e$ and $g_{NL}
\propto \e^2$
\cite{Koyama:2007if,Buchbinder:2007at,Lehners:2009ja}. In
addition, $f_{NL}$ is predicted to take a negative value and
$g_{NL}$ a positive one; the magnitude and sign do not fit well with current observations.

Finally, we note that, in spirit, our approach is somewhat
reminiscent of the $\d N$ formalism
\cite{Starobinsky:1986fxa,Sasaki:1995aw}. However, a full $\d
N$ calculation spanning both the generation and conversion of
the cosmological perturbations is made difficult here because
of the transition between the ekpyrotic and kinetic phases;
note, in particular, that in going from the ekpyrotic to the
kinetic phase, the equation of state drops drastically from
$w_{ek} \gg 1$ to $w_{kin} \approx 1$ (by contrast, the $\delta
N$ formalism is well adapted to new ekpyrotic models where the
generation and conversion both take place during the ekpyrotic
phase \cite{Koyama:2007if}). In fact, it is precisely the
disconnectedness between the two phases that allows our method
to work so well, and we would expect it to be applicable more
generally to cases where the physical properties of the phases
of generation and conversion differ substantially. The
separation between the two phases allows for a two-stage
approach in which we first solve for the entropy perturbation
during the ekpyrotic phase, and then use this as input for
calculating $\z$ or, equivalently $\d N,$ by perturbing around
the background trajectory during the kinetic phase. The simple,
yet non-perturbative, Eq. (\ref{Zetadot}) then reveals its full
effectiveness by yielding the result in just a few lines of
derivation. In this way, we have found a quick and rather
accurate way of understanding non-gaussianity in two-field
cyclic models of the universe.

\noindent {\it Acknowledgements} We would like to thank Justin Khoury for useful
discussions. This work is supported by US Department of Energy grant DE-FG02-91ER40671.




\end{document}